\documentstyle[epsfig]{mn1}
\newcommand{\da}{d_A}

\newcommand{\bp}{{\cal C}}    
\newcommand{\shell}{{\rm s}}     
\newcommand{\bfl}{{\mathbf{l}}} 
 \newcommand{\wj}{\left(
                          \begin{array}{ccc}
                          l_1  &  l_2  & l_3 \\
                            0  &  0    &  0
                          \end{array}
                          \right)}
\newcommand{\bi}{B_{l_1 l_2 l_3}}

\newcommand{\veck}{{\bf k}}
\newcommand{\vecl}{{\bf l}}
\newcommand{\bn}{{\bf \hat{n}}}

\newlength{\tskip}
\newlength{\colwidth}

\newcommand{\beq}{\begin{equation}}
\newcommand{\eeq}{\end{equation}}
\newcommand{\beqa}{\begin{eqnarray}}
\newcommand{\eeqa}{\end{eqnarray}}
\title[Second Order Corrections to Weak Lensing]{Second Order Corrections to Large Scale
Structure Weak  Lensing: Background Source Clustering}
\author[Cooray]{Asantha Cooray\\
Division of Physics, Mathematics and Astronomy, California
Institute of Technology, Pasadena, CA 91125.
E-mail: asante@caltech.edu}
\date{\today}
\begin{document}
\maketitle

%------------------------------------------------------------------------------

\begin{abstract}
We discuss the 
second order contributions to lensing statistics resulting from the
clustering of background sources from which galaxy shape measurements
are made in weak lensing experiments. In addition to a previously
discussed contribution to the lensing skewness, background source clustering
also contributes to the two-point correlation function, such as the
angular power spectrum of convergence or shear. 
At arcminute scales or above, the second order contribution to the
angular power spectrum of convergence due to source clustering is below the
level of a few percent. The background clustering of sources 
also results in a non-Gaussian contribution to
the power spectrum covariance of weak lensing convergence through a four-point
correlation function or a trispectrum in Fourier space.
The increase in variance is, at most, a few percent relative to the
Gaussian contribution while the band powers are also correlated at the
few percent level.  The non-Gaussian contributions due to background
source clustering is at least an order of magnitude smaller than those
resulting from non-Gaussian aspects of the large scale structure due to
the non-linear evolution of gravitational perturbations. We suggest that
the background source clustering is unlikely to affect the precision 
measurements of cosmology from upcoming weak lensing surveys.
\end{abstract}
 

\section{Introduction}

Weak gravitational lensing of faint galaxies probes the
distribution of matter along the line of sight.  Lensing by
large-scale structure (LSS) induces
correlation in the galaxy ellipticities at the few percent level
and can be detected through challenging statistical studies of galaxy shapes
in wide-field surveys (e.g., Blandford et al. 1991; Miralda-Escud\'e 1991;
Kaiser 1992).  An important aspect of weak lensing is that these ellipticity
correlations, or associated statistics,
 provide cosmological information that is, in some cases, complementary to those supplied by
the cosmic microwave background data at the same level of precision or better
(e.g., Jain \& Seljak 1997;
Bernardeau et al. 1997; Kaiser 1998; Schneider et al.
1998; Hu \& Tegmark 1999; Cooray 1999; van Waerbeke 1999;
see Mellier 1999 and Bartelmann \& Schneider 2000 for recent reviews).
Indeed, a wide number of recent studies have provided the clear evidence
for weak lensing due to large scale structure (van Waerkebe et al. 2000;
Bacon et al. 2000; Wittman et al. 2000; Kaiser et al. 2000),
though more work is clearly needed to understand even the statistical errors
and biases. While biases
 can result from certain aspects related to the selection of
observational fields (e.g., Cooray et al. 2000), statistical errors include
 those that are fundementally present due to the non-Gaussian nature of the
large scale structure (e.g., Cooray \& Hu 2001b).

Current predictions on statistics related to
large scale structure weak lensing  and
their ability to measure cosmological  parameters are based on several
assumptions; for example, it is implicitly assumed that background galaxies,
from which galaxy shape measurements are made, are uniformly distributed.
Another assumption is the use of so-called Born approximation where 
one integrates along unperturbed photon geodesics instead of perturbed photon paths.
Here, we discuss the former assumption while Cooray \& Hu (2002) presents a
discussion of the corrections resulting from dropping the Born approximation and
including the so-called lens-lens coupling between two lenses at two different redshifts.

The effect of background source clustering was first discussed with respect to the lensing
three point statistics, such as convergence skewness (Bernardeau 1998).
Hamana et al. (2001) includes an extended discussion of this
contribution. Recently, the same effect was revisited by
Schneider et al. (2002) as a possible source of curl-like modes in
lensing statistics. Here, we present a general discussion of the corrections resulting from
clustering of background sources and consider effects  two, three and four-point
correlation functions.  We suggest that the contributions are negligible
and are unlikely to affect the current and upcoming weak lensing observations
as a probe of cosmological parameters by comparing to predictions that neglect background
source clustering.

\section{Calculational Method}

We discuss our statistics in terms of lensing convergence, 
\begin{equation}
\kappa(\bn) = \frac{3}{2}\Omega_m H_0^2 \int_0^{\chi_0} d\chi
n_s(\chi) \int_0^{\chi} d\chi' g(\chi',\chi) \delta(\chi' \bn; \chi') \, ,
\end{equation}
where $n_s(\chi)$ is the normalized radial distribution of background
sources  such that $\int d\chi n_s(\chi)=1$ and $g(\chi',\chi)$ is the
lensing weight function when a background source is at a radial distance
of $\chi$:
\begin{equation}
g(\chi',\chi) = \frac{\da(\chi')\da(\chi-\chi')}{\da(chi)} \, .
\end{equation}
Here,  $\chi$ is the radial distance, or lookback time, 
from the observer, given by
\begin{equation}
\chi(z) = \int_0^z {dz' \over H(z')} \,,
\end{equation}
and the analogous angular diameter distance
\begin{equation}
\da(\chi) = H_0^{-1} \Omega_K^{-1/2} \sinh (H_0 \Omega_K^{1/2}
\chi)\,,
\end{equation}
with the expansion rate for adiabatic CDM cosmological models with a
cosmological constant given by
\begin{equation}
H^2 = H_0^2 \left[ \Omega_m(1+z)^3 + \Omega_K (1+z)^2
+\Omega_\Lambda \right]\,.
\end{equation}
Here, $H_0$ can be written as the inverse
Hubble distance today $cH_0^{-1} = 2997.9h^{-1} $Mpc.
We follow the conventions that in units of the critical
density $3H_0^2/8\pi G$,
the contribution of each component is denoted
$\Omega_i$, $i=c$ for the CDM, $b$ for the baryons,
$\Lambda$ for the cosmological constant.
We also define the auxiliary quantities
$\Omega_m=\Omega_c+\Omega_b$ and$\Omega_K=1-\sum_i \Omega_i$,
which represent the matter density and the contribution of
spatial curvature to the expansion rate respectively.
Note that as $\Omega_K \rightarrow 0$,
$\da(\chi) \rightarrow \chi$ and we define $\chi(z=\infty)=\chi_0$.
Though we present a general derivation of second order contributions
to weak lensing due to background source clustering, 
we show results for the currently
favorable $\Lambda$CDM cosmology with
$\Omega_b=0.05$, $\Omega_m=0.35$, 
$\Omega_\Lambda=0.65$ and $h=0.65$.

We assume that background sources are clustered in radial space such that
$n_s(\chi) \approx \bar{n}_s(\chi) [1+ \delta n_s(\chi)]$ where
fluctuations in the number counts of background sources are related to
that of the density field via a time, and possibly a scale dependent,
bias
\begin{equation}
\delta n_s(\chi) = b_1(\chi) \delta(\chi)+\frac{1}{2}b_2(\chi)
\delta^2(\chi) \, .
\end{equation}
Here we have considered, perturbatively, contributions up to the
second order in density perturbations. It is important that we
consider contributions up to the $\delta^2$ term since these terms can
contribute at the same second order level in the power spectrum. 
Though we discuss the effect in terms of 
lensing convergence, it should be noted that 
our calculation equally well apply for, correlations of shear, 
such as the gradient, or electric-like, modes (see, Schneider et al. 2002).

In Fourier space, we can write the first, second and third order contribution 
to convergence as
\begin{eqnarray}
&& \quad \kappa^{(1)}(\vecl) =  \frac{3}{2}\Omega_m H_0^2 \int d^2\theta
e^{-\vecl \cdot {\bf \theta}} \int_0^{\chi_0} d\chi
\bar{n}_s(\chi) \nonumber \\
&\times& \int_0^{\chi}  d\chi' g(\chi',\chi)  \int \frac{d^3\veck}{(2\pi)^3}
\delta(\veck,\chi')  e^{i \veck \cdot \chi' {\bf \theta}}  \, ,  \nonumber \\
&& \quad \kappa^{(2)}(\vecl) =  \frac{3}{2}\Omega_m H_0^2 \int d^2\theta
e^{-\vecl \cdot {\bf \theta}} \int_0^{\chi_0} d\chi
\bar{n}_s(\chi) b_1(\chi) \nonumber \\
&\times& \int \frac{d^3\veck_1}{(2\pi)^3}
\delta(\veck_1,\chi)  \int_0^{\chi} d\chi' g(\chi',\chi) \int \frac{d^3\veck_2}{(2\pi)^3}
\delta(\veck_2,\chi')  \nonumber \\
&& e^{i {\bf \theta} \cdot (\veck_1  \chi+\veck_2  \chi')} 
\nonumber \\
\end{eqnarray}
and
\begin{eqnarray}
&& \quad \kappa^{(3)}(\vecl) =  \frac{3}{4}\Omega_m H_0^2 \int d^2\theta
e^{-\vecl \cdot {\bf \theta}} \int_0^{\chi_0} d\chi
\bar{n}_s(\chi) b_2(\chi) \nonumber \\
&\times& \int \frac{d^3\veck_1}{(2\pi)^3}
\delta(\veck_1,\chi)  \int \frac{d^3\veck_2}{(2\pi)^3}
\delta(\veck_2,\chi)  
\int_0^{\chi} d\chi' g(\chi',\chi) \nonumber \\
&\times& \int \frac{d^3\veck_3}{(2\pi)^3}
\delta(\veck_3,\chi') 
 e^{i {\bf \theta} \cdot (\veck_1  \chi+\veck_2  \chi +\veck_3 \chi')}  \, ,
\nonumber \\
\end{eqnarray}
respectively.

We define the associated angular power spectrum, bispectrum and
trispectrum of weak lensing convergence, in flat-sky as appropriate
for current and upcoming experiments as
\begin{eqnarray}
\langle \kappa(\vecl) \kappa(\vecl') \rangle &=& (2\pi)^2
\delta_D(\vecl+\vecl') C_l^{\kappa \kappa} \nonumber \\
\langle \kappa(\vecl) \kappa(\vecl') \kappa(\vecl'')\rangle &=& (2\pi)^2
\delta_D(\vecl+...+\vecl'') B_\kappa(\vecl,\vecl',\vecl'') \nonumber
\\
\langle \kappa(\vecl) \kappa(\vecl') \kappa(\vecl'') \kappa(\vecl''')\rangle_c &=& (2\pi)^2
\delta_D(\vecl+...+\vecl''')
T_\kappa(\vecl,\vecl',\vecl'',\vecl''') \, . \nonumber \\
\end{eqnarray}
Here, $\veck_{i\ldots j} = \veck_i + \ldots + \veck_j$ and $\delta_D$ is
the delta function not to be confused with the density perturbation.
Note that the subscript $c$ denotes the connected piece, i.e., the
trispectrum is defined to be identically zero for a Gaussian field.
Here and throughout, we occasionally suppress the redshift dependence
where no confusion will arise.

We define the power spectrum of density fluctuations as
\begin{equation}
\langle \delta(\veck) \delta(\veck') \rangle = (2\pi)^3
\delta(\veck+\veck') P_{\delta \delta}(k) \, ,
\end{equation}
where
\begin{equation}
\frac{k^3P(k)}{2\pi^2} = \delta_H^2
\left({k \over H_0} \right)^{n+3}T^2(k) \,,
\end{equation}
in linear perturbation theory.
We use the fitting formulae of Eisenstein \& Hu (1999)
in evaluating the transfer function $T(k)$ for CDM models.
Here, $\delta_H$ is the amplitude of present-day density fluctuations
at the Hubble scale; with $n=1$, we adopt the COBE normalization
for $\delta_H$  (Bunn \& White 1997) of $4.2 \times 10^{-5}$,
consistent with galaxy cluster abundance (Viana \& Liddle 1999),
with $\sigma_8=0.86$. To capture the non-linear aspects of the power spectrum, 
we use the prescription by Peacock \& Dodds (1996). 

We will now discuss contributions 
to the angular power spectrum, bispectrum and trispectrum of convergence from
clustering of background sources.
To illustrate results, we
take a redshift distribution for the background  sources of the form
\begin{equation}
n_s(z) = \left(\frac{z}{z_0}\right)^{\alpha} \exp
-\left(\frac{z}{z_0}\right)^\beta \, ,
\end{equation}
where $(\alpha,\beta)$ denote the slope of the distribution at
low and high $z$'s, respectively around a mean-like
parameter given by $\sim z_0$.
For the purpose of this calculation we take $\alpha=\beta=1.5$ and
take $z_0$ to be 1.0 so as to mimic the expected background
sources from current and upcoming lensing catalogs. In general, since
clustering evolves to low redshifts, with a decrease in the background
source redshift distribution to a lower redshift, 
we expect an increase in the importance of second order effects
associated with clustering of background sources.

\subsection{Power Spectrum}

To the first order, using 
$\langle \kappa^{(1)}(\vecl) \kappa^{(1)}(\vecl')\rangle$, we simplify with
the Limber approximation (Limber 1954) in Fourier
space following Kaiser (1992; Kaiser 1998) to obtain the
well known result that
\begin{eqnarray}
&&C_l^{\kappa \kappa} = \frac{9}{4}\Omega_m^2 H_0^4 
\left[\int_0^{\chi_0} d\chi \bar{n}_s(\chi)\right]^2 
\nonumber \\
&\times& \int_0^{\chi} d\chi' \frac{g^2(\chi',\chi)}{\da(\chi')^2}
P_{\Phi\Phi}\left(\frac{l}{\da(\chi')};\chi'\right) \, . \nonumber \\
\end{eqnarray}

The second order convergence power spectrum resulting from source 
clustering involves
two terms: $\langle \kappa^{(2)} \kappa^{(2)} \rangle$
and $\langle \kappa^{(3)} \kappa^{(1)}\rangle$. The latter includes 
an additional, and equal, term through a  permutation.
We simplify these contributions 
again using the Limber approximation (Limber 1954)
such that
\begin{eqnarray}
&&\langle \kappa^{(2)}(\vecl) \kappa^{(2)}(\vecl') \rangle = 
(2\pi)^2
\delta_D(\vecl+\vecl') 
\nonumber \\
&\times& \frac{9}{4}\Omega_m^2 H_0^4  \int_0^{\chi_0} d\chi
\frac{[b_1 \bar{n}_s]^2(\chi)}{\da^2(\chi)} \int_0^{\chi}  d\chi' \frac{g^2(\chi',\chi)}{\da^2(\chi')}  \nonumber \\
&\times& \int \frac{d^2\vecl_1}{(2\pi)^2}
P\left(\frac{l_1}{\da(\chi)},\chi\right)
P\left(\frac{|\vecl-\vecl_1|}{\da(\chi')},\chi'\right) \, ,
\end{eqnarray}
and
\begin{eqnarray}
&&\langle \kappa^{(3)}(\vecl) \kappa^{(1)}(\vecl') \rangle = 
(2\pi)^2
\delta_D(\vecl+\vecl') \nonumber \\
&\times&
\frac{9}{8}\Omega_m^2 H_0^4 \int_0^{\chi_0} d\chi
\frac{b_2 \bar{n}^2_s(\chi)}{\da^2(\chi)} 
\int_0^{\chi}  d\chi' \frac{g^2(\chi',\chi)}{\da^2(\chi')}  
\nonumber \\
&\times& 
\int \frac{d^2\vecl_1}{(2\pi)^2}
P\left(\frac{l_1}{\da(\chi)},\chi\right)
P\left(\frac{l}{\da(\chi')},\chi'\right) \, . \nonumber \\
\end{eqnarray}
Note that the total contribution to the convergence power spectrum 
follows as $\langle \kappa^{(2)}(\vecl) \kappa^{(2)}(\vecl') \rangle$+
$\langle \kappa^{(3)}(\vecl) \kappa^{(1)}(\vecl') \rangle$+
$\langle \kappa^{(1)}(\vecl) \kappa^{(3)}(\vecl') \rangle$.
We denote the first contribution by $C_l^{22}$ term and the latter two
terms by $C_l^{31}$. Thus, the total contribution to the power
spectrum due to source clustering is
\begin{eqnarray}
C_l^{\kappa\kappa} &=& C_l^{22} + C_l^{31} \nonumber \\
C_l^{22} &=& \frac{9}{4}\Omega_m^2 H_0^4  \int_0^{\chi_0} d\chi
\frac{[b_1 \bar{n}_s]^2(\chi)}{\da^2(\chi)} \int_0^{\chi}  d\chi' \frac{g^2(\chi',\chi)}{\da^2(\chi')}  \nonumber \\
&\times& \int \frac{d^2\vecl_1}{(2\pi)^2}
P\left(\frac{l_1}{\da(\chi)},\chi\right)
P\left(\frac{|\vecl-\vecl_1|}{\da(\chi')},\chi'\right) \nonumber \\
C_l^{31} &=& \frac{9}{4}\Omega_m^2 H_0^4 \int_0^{\chi_0} d\chi
\frac{b_2 \bar{n}^2_s(\chi)}{\da^2(\chi)} 
\int_0^{\chi}  d\chi' \frac{g^2(\chi',\chi)}{\da^2(\chi')}  
\nonumber \\
&\times& 
\int \frac{d^2\vecl_1}{(2\pi)^2}
P\left(\frac{l_1}{\da(\chi)},\chi\right)
P\left(\frac{l}{\da(\chi')},\chi'\right) \, . \nonumber \\
\end{eqnarray}

\begin{figure}
\psfig{file=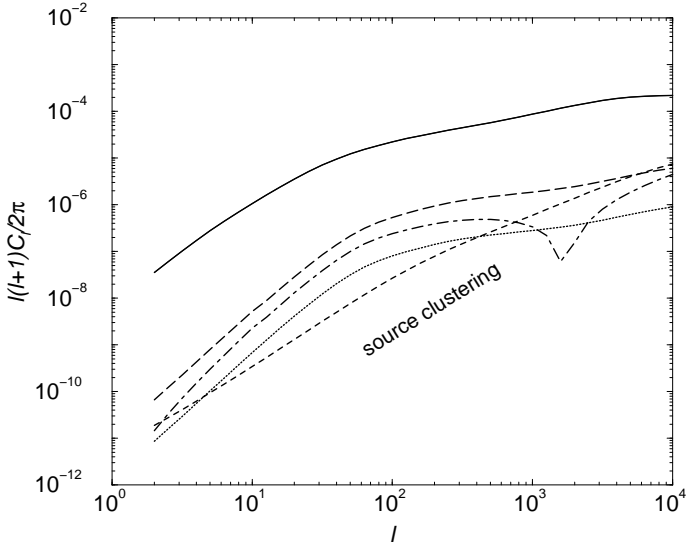,width=2.9in,angle=-90}
\caption{The angular power spectrum of convergence. The solid line
shows the first order contribution while corrections due to background
source clustering is shown with dashed ($C_l^{22}$) and long-dashed
($C_l^{31}$) lines. We have assumed $b_1=1$ and $b_2=1$ in these two
cases, respectively. The dot-dashed line is the total second order
power spectrum when $b_1=1$ and $b_2=-0.5$ consistent with suggestions
in the literature for galaxy bias (see text for details). The dotted line is 
the
$C_l^{31}$ contribution when galaxies trace the linear density field instead of the fully non-linear power spectrum.}
\label{fig:cl}
\end{figure}

In figure~\ref{fig:cl}, we show the second order correction to the
angular power spectrum of convergence.   The solid line is the well
known first order result, while the dashed line is the contribution
from the $C_l^{22}$ term. Note that
$C_l^{22} \propto b_1^2$, and we have taken the value of $b_1=1$ for
illustration purposes. The  long-dashed line is the $C_l^{31}$ contribution,
where $C_l^{31} \propto b_2$ and we have taken $b_2=1$ for simplicity. 
Since we do not have detailed information on the galaxy bias,
to illustrate our results, 
we have taken the bias to be redshift and scale independent. 
Adding a redshift dependent 
bias of the from $b(z) \propto (1+z)^\gamma$, however, did
not lead to a significantly different result from the one 
suggested in figure~\ref{fig:cl} when $|\gamma| <$ 2.

There is one aspect of bias that should be kept in
mind when interpreting figure~\ref{fig:cl}. 
In general, quadratic bias is expected to be negative, such that the two terms,
$C_l^{22}$ and $C_l^{31}$, added 
together will give a contribution which is lower than what one would
naively expect if simply added together. In the IRAS PSCz catalog,
Feldman et al. (2001) finds $1/b_1 = 1.20^{+0.18}_{-0.19}$ and
$b_2/b_1^2 = -0.42 \pm 0.19$. Similar results our available
from the 2dF survey by Verde et al. (2002): $b_1=1.04 \pm 0.11$
and $b_2=-0.054\pm0.08$. Since there is no conclusive evidence for a
non-zero value for $b_2$, the first approximation that $b_2=0$
and $b_1=1$ leads to the dashed 
line with the conclusion that second order effects are
generally below the few percent level for multipole less than $10^{4}$
corresponding to angular scales less than few arcminutes.

On the other hand, if $b_2=-0.5$ and $b_1=1$, we obtain the dot-dashed line
as the total contribution to the angular power spectrum of convergence
due to source clustering; the shape of the power spectrum is due to
the fine cancellation of $C_l^{22}$ and $C_l^{31}$ terms.
We suggest that, in addition to the linear bias, the extent
to which background source clustering affects statistics such as power
spectra or correlations depends on the detail aspects of
galaxy biasing such as the quadratic bias.
In any case, we find that source clustering effects are unlikely to be a 
strong contaminat for current lensing experiments.

There is also another important aspect related to $C_l^{31}$. The integral over 
$\vecl_1$  denotes the power spectrum traced by galaxies and $C_l^{31}$ effectively
scales with this integral as an overall normalization. If galaxies do not fully
trace the non-linear power spectrum, as predicted by the Peacock \& Dodds (1996) formulae
for the dark matter,
 then the contribution would be lower than what we have predicted.
We can bracket the expected range of variation by replacing the power 
spectrum involved with the
integral over $\vecl_1$ with that of the linear power spectrum, as it is 
generally expected that
the galaxy power spectrum lies between the linear and non-linear cases of 
dark matter.
Since the contribution to the integral here comes from all angular scales, the bahvior of either
the linear or non-linear power spectrum
 at small angular scales becomes to some extent important for the
calculation presented here. As we do not have a reliable method to predict the non-linear power
spectrum at small scales, we safely cut off the calculation at $k \sim 10^{6}$ h Mpc$^{-1}$.

\subsection{Bispectrum}

We can write the resuling contribution to the bispectrum by
considering terms such as $\langle \kappa^{(2)}(\vecl_1)
\kappa^{(1)}(\vecl_2) \kappa^{(1)}(\vecl_3) \rangle$ and add the
necessary permutations. We write one of these terms as
\begin{eqnarray}
&&\langle \kappa^{(2)}(\vecl) \kappa^{(1)}(\vecl')
\kappa^{(1)}(\vecl'') \rangle = 
(2\pi)^2 \delta_D(\vecl+\vecl'+\vecl'') 
\left(\frac{3}{2}\Omega_m H_0^2\right)^3  
\nonumber \\ &\times&
\int_0^{\chi_0} d\chi
\bar{n}_s(\chi) \int_0^{\chi}  d\chi' \frac{g(\chi',\chi) b_1(\chi') \bar{n}_s(\chi')}{\da^2(\chi')}  
P\left(\frac{l_1}{\da(\chi')},\chi'\right) \nonumber \\
&\times&\int_0^{\chi_0} d\chi
\bar{n}_s(\chi) \int_0^{\chi}  d\chi' \frac{g^2(\chi',\chi)}{\da^2(\chi')}  
P\left(\frac{l_2}{\da(\chi')},\chi'\right) \, ,\nonumber \\
\end{eqnarray}
so that the bispectrum is
\begin{eqnarray}
&& B_\kappa(\vecl_1,\vecl_2,\vecl_3) =
\left(\frac{3}{2}\Omega_m H_0^2\right)^3  
\nonumber \\ &\times&
\int_0^{\chi_0} d\chi
\bar{n}_s(\chi) \int_0^{\chi}  d\chi' \frac{g(\chi',\chi) b_1(\chi') \bar{n}_s(\chi')}{\da^2(\chi')}  
P\left(\frac{l_1}{\da(\chi')},\chi'\right) \nonumber \\
&\times&\int_0^{\chi_0} d\chi
\bar{n}_s(\chi) \int_0^{\chi}  d\chi' \frac{g^2(\chi',\chi)}{\da^2(\chi')}  
P\left(\frac{l_2}{\da(\chi')},\chi'\right) +{\rm Perm.} \, ,\nonumber \\
\end{eqnarray}
where the permutations are with respect to the ordering of
$(l_1,l_2,l_3)$ and involves five additional terms.

Following Cooray \& Hu (2001a), we can write the third 
moment of convergence using
the bispectrum as 
\begin{eqnarray}
\left< \kappa^3(\sigma) \right> &=&
                {1 \over 4\pi} \sum_{l_1 l_2 l_3}
                \sqrt{(2l_1+1)(2l_2+1)(2l_3+1) \over 4\pi} \nonumber\\
                &&\times \wj \bi^\kappa
                W_{l_1}(\sigma)W_{l_2}(\sigma)W_{l_3}(\sigma)
                \,, \nonumber \\
\end{eqnarray}
where the quantity within parantheses is the Wigner-3$j$ symbol, which
in the case of no angular dependence can be written as
\begin{eqnarray}
\wj &=&
(-1)^{L/2} \frac{({L \over 2})!}{({L \over 2}-l_1)!({L \over 2}-l_2)!({L \over 2}-l_3)!}
\nonumber \\
&\times&
\left[\frac{(L-2l_1)!(L-2l_2)!(L-2l_3)!}{(L+1)!}\right]^{1/2}
\nonumber \\
\end{eqnarray}
for even $L$; it vanishes for odd $L$. We refer the reader to
Cooray \& Hu (2000) for additional details on the Wigner 3-$j$ symbol.

Using the third moment, we construct the skewness as
\begin{equation}
S_3(\sigma) =
\frac{\left<\kappa^3(\sigma)\right>}{\left<\kappa^2(\sigma)\right>^2}
\, .
\end{equation}
where the all-sky expression for bispectrum, in terms of the flat-sky
derivation, is
\begin{eqnarray}
\bi^\kappa &=& \sqrt{(2l_1+1)(2l_2+1)(2l_3+1)  \over 4\pi} \wj
        \nonumber\\
&&\times B_\kappa(l_1,l_2,l_3) \, .
\label{eqn:szbispectrum}
\end{eqnarray}
Similarly, the second moment is defined as
\begin{equation}
\left< \kappa^2(\sigma) \right> =
{1 \over 4\pi} \sum_l (2l+1) C_l^\kappa W_l^2(\sigma)\,.
\label{eqn:secondmom}
\end{equation}
where $W_l$ are the multipole moments (or Fourier transform in a
flat-sky approximation) of the window.   For simplicity, we will
choose
a window which is a two-dimensional top hat in real space with a
window function in
multipole space of $W_l(\sigma) = 2J_1(x)/x$ with $x = l\sigma$.

\begin{figure}
\psfig{file=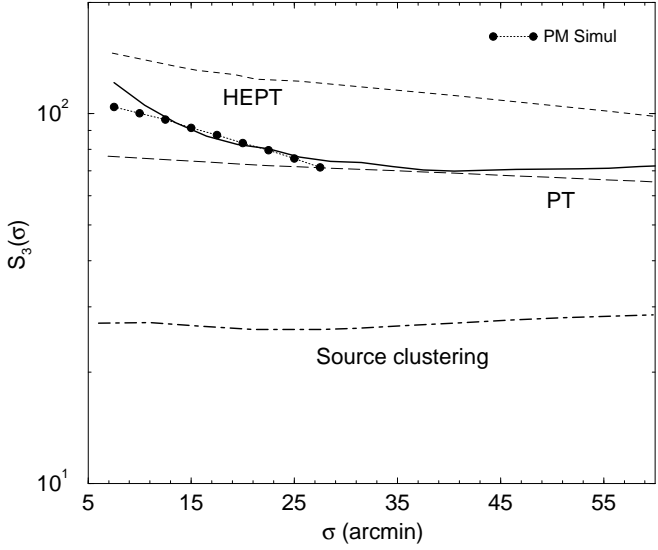,width=2.9in,angle=-90}
\caption{The skewness of convergence. We show the results based on
gravitational perturbation theory calculation to the second order
(PT; dashed line), and the calculation valid for highly non-linear regime based on
hyper-extended perturbation theory (HEPT; Scoccimarro \& Frieman 1999; dotted
line). The solid line is the expected skewness based on the halo model
calculation following Cooray \& Hu (2001). For comparison, we show
results from N-body particle-mesh simulations by White \& Hu (1999).
The higher order correction to skewness resulting from background
source clustering is shown with a dot-dashed line. The second order
contribution is at the level of few tens of percent of the total
expected from non-linear evolution of gravitational perturbations.}
\label{fig:skew}
\end{figure}

In figure~\ref{fig:skew}, 
we summarize our results on the expected contribution to
the convergence
skewness. We refer the reader to Bernardeau (1998) and Hamana et al. (2002) for
an extended discussion on the effects on background source clustering
on skewness. As shown in figure~\ref{fig:skew} and discussed
in prior publications, the contribution to skewness is at the level of
few tens of percent and depends strongly on parameters such as the mean redshift of
background sources and the width of the redshift distribution. A higher mean redshift and
a smaller width result in  a smaller contribution to skewness while
a lower mean redshift and a broader distribution can contribute
 up to 30\% or more 
of
the skewness expected from non-linear evolution of gravitational perturbations.

\begin{figure}
\psfig{file=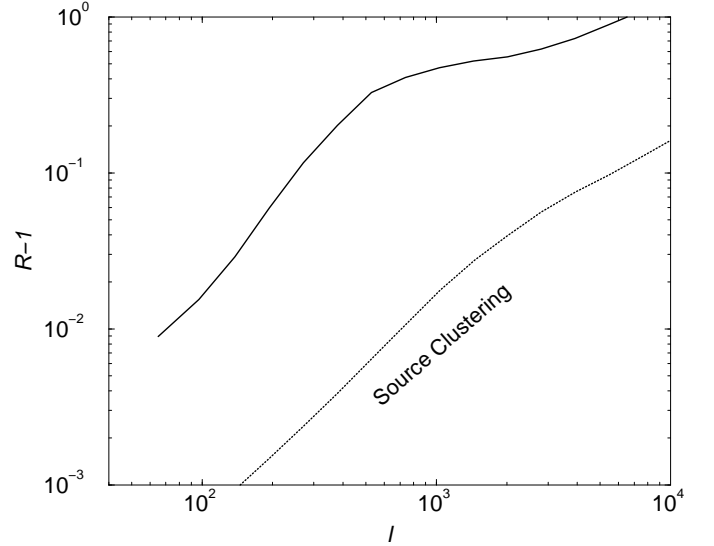,width=2.9in,angle=-90}
\caption{The ratio of non-Gaussian to Gaussian error. The solid line
is the calculation for non-linear evolution of gravitational
perturbations using dark matter halos following the prescription in
Cooray \& Hu (2001b), while the dotted line is contribution arising
from source clustering effects. We use the same binning scheme in
multipole space as Cooray \& Hu (2001b).}
\label{fig:tri}
\end{figure}

\subsection{Trispectrum}

We can write the resuling contribution to the trispectrum by
considering terms such as $\langle \kappa^{(2)}(\vecl_1)
\kappa^{(2)}(\vecl_2) \kappa^{(1)}(\vecl_3) \kappa^{(1)}(\vecl_4) \rangle$ and add the
necessary permutations. We write one of these terms as
\begin{eqnarray}
&&\langle \kappa^{(2)}(\vecl) \kappa^{(2)}(\vecl')
\kappa^{(1)}(\vecl'') \kappa^{(1)}(\vecl''') \rangle = \nonumber \\
&&(2\pi)^2 \delta_D(\vecl+\vecl'+\vecl''+\vecl''') 
\left(\frac{3}{2}\Omega_m H_0^2\right)^4 
\nonumber \\ 
&\times& 
\int_0^{\chi_0} d\chi_1
\frac{[b_1\bar{n}_s]^2(\chi_1)}{\da^2(\chi_1)} \left[
P\left(\frac{|\vecl_3+\vecl_1|}{\da(\chi_1)},\chi_1\right) +
P\left(\frac{|\vecl_3+\vecl_2|}{\da(\chi_1)},\chi_1\right) \right] \, ,\nonumber \\
&\times& \int_0^{\chi_0} d\chi
\bar{n}_s(\chi) \int_0^{\chi_2}  d\chi' \frac{g(\chi',\chi_1)g(\chi',\chi)}{\da^2(\chi')}  
P\left(\frac{l_1}{\da(\chi')},\chi'\right) \nonumber \\
&\times&\int_0^{\chi_0} d\chi
\bar{n}_s(\chi) \int_0^{\chi}  d\chi' \frac{g(\chi',\chi_1)g(\chi',\chi)}{\da^2(\chi')}  
P\left(\frac{l_2}{\da(\chi')},\chi'\right) \, , \nonumber \\
\end{eqnarray}
so that the trispectrum is
\begin{eqnarray}
&& T_\kappa(\vecl_1,\vecl_2,\vecl_3,\vecl_4) = \left(\frac{3}{2}\Omega_m H_0^2\right)^4 \nonumber \\
&\times& \int_0^{\chi_0} d\chi_1
\frac{[b_1\bar{n}_s]^2(\chi_1)}{\da^2(\chi_1)} \left[
P\left(\frac{|\vecl_3+\vecl_1|}{\da(\chi_1)},\chi_1\right) +
P\left(\frac{|\vecl_3+\vecl_2|}{\da(\chi_1)},\chi_1\right) \right] \, ,\nonumber \\
&\times& \int_0^{\chi_0} d\chi
\bar{n}_s(\chi) \int_0^{\chi_2}  d\chi' \frac{g(\chi',\chi_1)g(\chi',\chi)}{\da^2(\chi')}  
P\left(\frac{l_1}{\da(\chi')},\chi'\right) \nonumber \\
&\times&\int_0^{\chi_0} d\chi
\bar{n}_s(\chi) \int_0^{\chi}  d\chi' \frac{g(\chi',\chi_1)g(\chi',\chi)}{\da^2(\chi')}  
P\left(\frac{l_2}{\da(\chi')},\chi'\right) \, , \nonumber \\
\end{eqnarray}
where the permutations are with respect to the ordering of
$(l_1,l_2,l_3,l_4)$.

For the purpose of this calculation, we assume that upcoming weak
lensing convergence power spectrum will measure binned logarithmic
band powers at several $l_i$'s in multipole space with bins of
thickness $\delta l_i$.
\begin{equation}
\bp_i =
\int_{\shell i}
{d^2 l \over{A_{\shell i}}}
\frac{l^2}{2\pi} \kappa(\bf l) \kappa(-\bf l) \, ,
\end{equation}
where $A_\shell(l_i) = \int d^2 l$ is the area of 2D shell in
multipole and can be written as $A_\shell(l_i) = 2 \pi l_i \delta l_i
+ \pi (\delta l_i)^2$.
 
We can now write the signal covariance matrix
as
\begin{eqnarray}
C_{ij} &=& {1 \over A} \left[ {(2\pi)^2 \over A_{\shell i}} 2 \bp_i^2
+ T^\kappa_{ij}\right]\,,\\
T^\kappa_{ij}&=&
\int {d^2 l_i \over A_{\shell i}}
\int {d^2 l_j \over A_{\shell j}} {l_i^2 l_j^2 \over (2\pi)^2}
T^\kappa(\bfl_i,-\bfl_i,\bfl_j,-\bfl_j)\,,
\label{eqn:variance}
\end{eqnarray}
where
$A$ is the area of the survey in steradians.  Again the first
term is the Gaussian contribution to the sample variance and the
second the non-Gaussian contribution.
A realistic survey will also have shot noise variance due to
the finite number of source galaxies in the survey.
For a comparison of previous calculations, we take the same binning
scheme as the one used in Cooray \& Hu (2001b) and used in
White \& Hu (1999).

In figure~\ref{fig:tri}, we show 
the ratio of $R \equiv C_{ii}/C_{ii}^{\rm Gaus}$ where
$C_{ii}^{\rm Gaus}$ is the contribution with simply the Gaussian variance.
This ratio can also be written as
\begin{equation}
R \equiv 1 + \frac{A_{si}T^\kappa_{ii}}{(2\pi)^2 2 \bp_i^2} \, ,
\end{equation}
and we plot $R-1$ to highlight the difference between source clustering and non-Gaussian aspect of large scale
structure. As shown, the clustering only leads to a
few percent contribution, at $l \sim 10^3$, beyond the Gaussian 
variance while the
non-Gaussianities due to large scale structure clustering contributes at the 
level of 10\% or more.
One can safely ignore the relative increase in the variance of power spectrum measurements due to
background source clustering.

\begin{figure}
\psfig{file=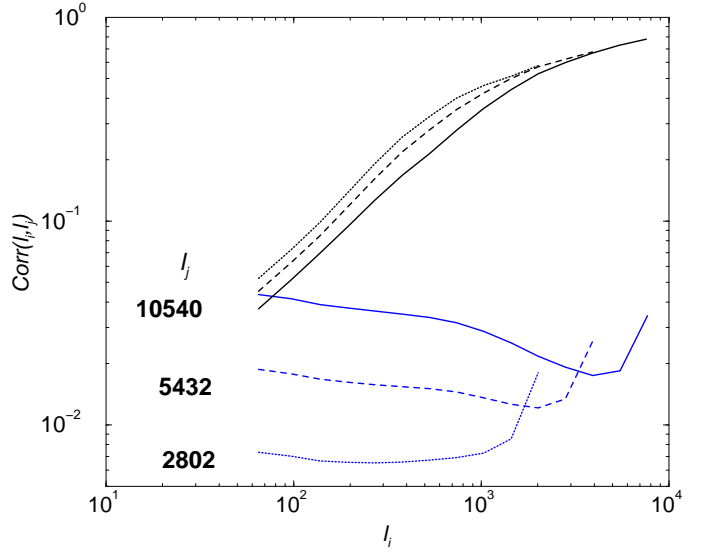,width=2.9in,angle=-90}
\caption{The correlation coefficient $\hat{C}_ij$, as a function of the multipole $l_i$ with $l_j$, as shown in the figure. The top lines are the
correlations due to the non-linear evolution of gravitational perturbations,
as calculated under the halo approach of Cooray \& Hu (2001b).
The bottom lines are the resulting correlations due to the
clustering of background  sources from which shape measurements are made.
These correlations are at the level of few percent, while non-Gaussian nature of
the large scale structure induces correlations at the tens of percent level or more.}
\label{fig:corr}
\end{figure}

An additional aspect of the covariance resulting from non-Gaussianities is that band power estimates are correlated.
These correlations can be written as
\begin{equation}
\hat C_{ij} \equiv \frac{C_{ij}}{\sqrt{C_{ii} C_{jj}}} \, .
\end{equation}
In figure~\ref{fig:corr},
we show the behavior of the correlation
coefficient between a fixed $l_j$ as a function of three $l_i$'s.  
When $l_i=l_j$
the coefficient is 1 by definition.  Due to the presence of
the dominant Gaussian contribution at $l_i=l_j$, the coefficient has 
an apparent
discontinuity between $l_i=l_j$ and $l_i = l_{j-1}$ that decreases
as $l_j$ increases and non-Gaussian effects dominate.
As shown, however, the correlation coefficients due to the
non-Gaussian nature of the large scale structure is over an order of
 magnitude larger than
than the correlations resulting from the clustering of background sources.
The results related to the covariance suggests that non-Gaussian effects
resulting from the clustering of background sources is not expected to
strongly influence the abilities of weak lensing experiment to obtain precision measurements of
cosmology.

\section{Summary \& Conclusions}

We have discussed the second order contributions to weak gravitational lensing
convergence resulting from the clustering of background sources from which galaxy shape measurements
are made in weak lensing experiments. The clustering of source galaxies induce a 
second order contribution to the two point statistics such as weak lensing
convergence angular power spectrum. For the angular scales of interest, we 
have shown that this
contribution is at the level of few percent; to some extent, however, the 
exact contribution is
uncertain due to unknown aspects associated with galaxy biasing such as the 
quadractic bias. 

Our calculations related to the skewness generated by background source 
clustering is
consistent with previous calculations by Bernardeau (1998) and 
Hamana et al. (2001).
We have discussed a new non-Gaussian aspect of the background source 
clustering involving a contribution to
the four point correlation function of shear or the trispectrum in Fourier 
space.
The non-Gaussian four-point function is of interest since it determines the 
covariance of power
spectrum measurements. The background source clustering increases the Gaussian covariance at the
level of few percent when $l \sim 10^3$. This increase, however, is an order of magnitude or more below the
increase resulting from the  intrinsic non-Gaussian nature of the large scale structure due to the non-linear
evolution of gravitational perturbations. The trispectrum contribution to the covariance also
leads to correlations between band power estimates, though, these are again at the
few percent level or below and are unlikely to be a significant source of error  for current
and upcoming weak lensing experiments.

\section*{Acknowledgments}
 
We thank support from the Sherman Fairchild foundation and the
Department of Energy grant DE-FG03-92-ER40701.

\end{document}